\documentclass[%
 reprint,
 amsmath,amssymb,
 aps,
onecolumn]{revtex4-2}

\usepackage{tikz}

\usepackage{graphicx}
\usepackage{dcolumn}
\usepackage{bm}
\usepackage{amsfonts}
\usepackage{amsmath}
\usepackage{float}
\usepackage{caption}
\usepackage{color}
\usepackage{physics}
\def\D{\mathrm{d}}
\def\I{\mathrm{i}}
\def\E{\mathrm{e}}
\usepackage{bbm}
\usepackage[bb=boondox]{mathalfa}

\newcommand{\vv}[1]{\boldsymbol{#1}}

\begin{document}

\preprint{APS/123-QED}

\title{Time Varying Resonators in Acoustic Waveguides: \\A Transfer Matrix Formalism for Space-time Modulated Metamaterials}

\author{
D. Cidlinsk\'{y}$^{1,2}$,}
\email{darek@mail.muni.cz}
\author{
G. J. Chaplain$^{2}$ and  S. A. R. Horsley$^{2}$
}

\address{$^{1}$Department of Theoretical Physics and Astrophysics, Faculty of Science, Masaryk University, Kotlářská 2, 61137 Brno, Czechia \\
$^{2}$Centre for Metamaterial Research and Innovation, Department of Physics and Astronomy, University of Exeter, United Kingdom}

\begin{abstract}
The transfer matrix method remains a simple yet powerful tool for modeling acoustic systems, particularly in a closed waveguide geometry. Here we present a generalisation of this method based on the theory of mode matching, that incorporates the effect of ultra--fast temporal variations in the geometry, applying it to a system of side-branching resonant cavities (quarter wavelength resonators) fixed to an acoustic waveguide, modulated through alteration of the cavity length.  We calculate propagation in a waveguide containing both a single resonator and a periodic array.  In particular we predict the generation of additional Doppler-like terms in the reflected and transmitted fields that leads to modification of the band structure, comparing our results to finite element simulations of space-time modulated acoustic crystals. 
\end{abstract}



\maketitle

\section{Introduction}

The concept of a metamaterial---that sub--wavelength structure can yield a zoo of otherwise impossible material properties~\cite{kadic2019}---has, since around the turn of the millennium, travelled far beyond its electromagnetic origins.  Acoustic~\cite{cummer2016}, elastic, thermal~\cite{li2021}, and water-wave~\cite{farhat2008,han2022} metamaterials have all been the subject of intense investigation.  Whereas interest to--date has largely concentrated on phenomena such as negative refraction~\cite{pendry2000}, and transformation optics--based designs such as invisibility cloaking~\cite{pendry2006,li2008}, as well as more recent work on topological materials~\cite{wang2009} and numerically optimized designs~\cite{hughes2018}, many researchers are turning their attention to the \emph{temporal} structure of the material response.  So called time--varying metamaterials~\cite{galiffi2022,engheta2023four,Lise1,Lise2} are externally controlled structures where the effective material parameters vary on a time scale on the order of the wave period.  Note that the metamaterial concept \emph{proper} rarely applies for current research in this area: we are usually not in the limit of ultra--fast modulation, where effective medium theory applies, although effective medium theory has been developed for time varying materials~\cite{pacheco2020,huidobro2021,zhu2023effective}.

Work on wave propagation in time--varying materials has been considered at least since the work of Morgenthaler in the 1950s~\cite{morgenthaler1958velocity}, with a few notable works since then (e.g.~\cite{mendonca2000,mendonca2002}).  This renewed attention is due to recent experimental progress.  For example, it has been realized that a family of materials (including Indium Tin Oxide) have an optical response that can be modulated with high contrast and over a short interval of time~\cite{zhou2020,tirole2023double,harwood2024super}.  Meanwhile, in acoustics, electronic feedback has been applied~\cite{wen2022} to realise effective time--varying properties in addition to a a highly controlled dispersive response.  Recent investigations have included the study of temporal boundary conditions \cite{xiao2014reflection}, wave propagation in temporally dispersive media \cite{solis2021time}, `temporal aiming'~\cite{pacheco2020b}, temporal anti--reflection coatings~\cite{pacheco2020c}, static-to-dynamic field conversion \cite{mencagli2022static}, `spoofing' motion induced effects \cite{huidobro2019fresnel,horsley2024traveling}, and spatio--temporal diffraction \cite{taravati2019generalized,tirole2023double,harwood2024super}. Several of these effects are yet to be observed.  Time--varying acoustic~\cite{zhu2023effective,chen2021efficient} and elastic~\cite{movchan2022frontal,liu2024inherent,wang2025temporal} materials offers a route to a simple experimental realisation, given the much slower wave speeds compared to the speed of light.

To--date, several methods have been developed for describing wave--propagation in time-varying materials (for instance, the operator method discussed in~ \cite{horsley2023eigenpulses}, developed to treat the combination of dispersion and time modulation).  Here our focus is on acoustic systems, where the analysis can be largely carried out analytically and with relative simplicity. In our case the time modulation of the effective medium response arises through a modulation of the material geometry in time.   We consider the simplest example of an acoustic quarter wavelength resonator in a waveguide: a parallel branch in a pipe. Models exist in which, by modulating either resonant frequency, line-width, or coupling strength temporally, effective material parameters can be derived resulting from the time-modulation \cite{zhu2023effective}. Despite this---to the best of our knowledge---there are no quick numerical schemes to quickly incorporate such time-modulated parameters in arrays of acoustic scatterers.

Here we present an extension of the familiar Transfer Matrix Method (TMM), that incorporates temporal variations in the resonators' resonant frequency to address this.  We derive and formalise, through modal matching of the pressure field at the mouth of the cavity, a time-varying TMM (TV-TMM) that incorporates, to first order, the additional sum and difference frequencies that originate due to the time variation; these can be viewed as Doppler-like terms (see e.g.~\cite{eden1992search}). We consider first a single time-dependent cavity, ignoring any sound generated by what in essence is a side-branching piston, before extending to infinitely periodic arrays and compare results to the classical TMM and with time-domain Finite Element (FE) simulations. 
\section{Formulation}
\label{sec:formulation}
%
\subsection{Scattering from a time-independent cavity\label{sec:time-independent}}
\par
We firstly briefly re-cap known results for a single time-independent cavity connected to a waveguide: the single cavity limit of the periodic array sketched in Fig. 1c.  The results we develop in this section will then seed the next section, where we extend the theory to the case of a modulated cavity connected to a waveguide.

For simplicity we take the waveguide and cavity to extend infinitely out of the page, allowing us to work in two dimensions, $x$, and $y$, where $x$ is the direction of propagation within the waveguide, and $y$ is the orthogonal direction, specifying the vertical distance from the lower surface of the waveguide. We fix the height of the waveguide equal to 1 measuring all other distances relative to this.  The cavity we take as width $2a$, and the depth of the cavity we take as fixed, $d(t) = d_0$.   Although we study the single cavity limit for now, we will use the results to extend to the periodic case, assuming a $2(L+a)$ spacing between the centres of the cavities. 

All of our calculations are performed in terms of the velocity potential $\phi$ ($\boldsymbol{v}=\boldsymbol{\nabla}\phi$), which, in the case of linear acoustics, satisfies the wave equation $\Delta\phi - c^{-2}\partial_t^2\phi = 0$, where $c$ is the speed of sound in air.  In the case of a time independent cavity, we can make the common assumption that the solution is of the form $\phi = \phi(x,y) \E^{-\I\omega t}$. Denoting $K = \omega/c$, we get the usual Helmholtz equation 
\begin{equation}
	(\Delta + K^2) \phi = 0 \quad \text{with} \quad \pdv{\phi}{\vv n} = 0 \quad \text{on all boundaries}.\label{eq:helm}
\end{equation}
where $\boldsymbol{n}$ is the unit vector normal to the boundary, implying a vanishing normal velocity on all of the boundaries.  If we confine our interest to low frequency waves within the system we can assume that only the fundamental, lowest order mode is supported in the cavity.  Within the cavity, we thus assume $\phi(x,y) = b \cos (K(y+d))$, with some complex amplitude $b$ so that $\phi$ satisfies the boundary conditions of zero gradient at the bottom of the cavity, $\partial_y\phi=0$, and on the cavity walls, $\partial_x\phi=0$.

Having given the solution to the Helmholtz equation (\ref{eq:helm}) within the adjoining cavity, we now seek the form of the wave in the waveguide above.  This is simplified if we write the velocity potential as a Fourier integral in the $x$ variable, $\phi(x,y) = (2\pi)^{-1} \int_{-\infty}^{\infty} \tilde \phi(y) \E^{\I kx}\,\D k$.  This choice leads to the following modified wave equation for each Fourier component $\tilde{\phi}$
\begin{equation}
    \frac{\partial^2\tilde{\phi}}{\partial y^2}+\left(K^2-k^2\right)\tilde{\phi}=0
\end{equation}
with the general solution
\begin{equation}
    \tilde{\phi}(k,y)=A(k)\cos(\sqrt{K^2-k^2}y)+B(k)\sin(\sqrt{K^2-k^2}y).
\end{equation}
To determine the functions $A(k)$ and $B(k)$, we now impose the boundary conditions on the upper and lower walls of the waveguide.  On the upper interface $y=1$ we have simply the condition for vanishing velocity $\partial_y\tilde{\phi}$, which is,
\begin{equation}
    \sqrt{K^2-k^2}\left[-A(k)\sin(\sqrt{K^2-k^2})+B(k)\cos(\sqrt{K^2-k^2})\right]=0,
\end{equation}
This condition allows us to eliminate the $A(k)$,
\begin{equation}
        A(k)=B(k)\cot(\sqrt{K^2-k^2})
    +2\pi\sum_{n=0}^{\lfloor K/\pi \rfloor}\left[c_n^{(+)}\delta\left(k-\sqrt{K^2-n^2\pi^2}\right)+c_n^{(-)}\delta\left(k+\sqrt{K^2-n^2\pi^2}\right)\right].\label{eq:Ak}
\end{equation}
where the $c_n^{(\pm)}$ represent the amplitudes of the different waves input from $x=\pm\infty$, which we can specify arbitrarily.  Meanwhile, on the lower surface ($y=0$), the normal derivative of the velocity potential must be zero everywhere, except at the opening of the cavity.  Taking the Fourier transform of the normal derivative on this lower interface then gives us the condition on the unknowns $B(k)$,
\begin{equation}
    B(k)=-\frac{Kb\sin(Kd)}{\sqrt{K^2-k^2}}\int_{-a}^{a}{\rm d} x \,{\rm e}^{-{\rm i}kx}=-\frac{2Kb\sin(Kd)\sin(ka)}{k\sqrt{K^2-k^2}}.\label{eq:Bk}
\end{equation}
In Eqns. (\ref{eq:Ak}) and (\ref{eq:Bk}) we have thus determined the form of the field in the waveguide, up to the constants $c_n^{(\pm)}$ which we are free to choose, and the complex amplitude $b$ of the field in the cavity, which we are yet to determine.  Combining these two results, the Fourier representation of the field $\tilde{\phi}(k,y)$ then equals
 \begin{multline}
 \tilde{\phi}(k,y)=-\frac{2Kb\sin(Kd)\sin(ka)}{k\sqrt{K^2-k^2}\sin(\sqrt{K^2-k^2})}\cos(\sqrt{K^2-k^2}(1-y))\\
 +\sum_{n=0}^{\lfloor K/\pi \rfloor}\bigg[c_n^{(+)}\delta\left(k-\sqrt{K^2-n^2\pi^2}\right)
 +c_n^{(-)}\delta\left(k+\sqrt{K^2-n^2\pi^2}\right)\bigg]\cos(\sqrt{K^2-k^2}y).
 \end{multline}
Performing the inverse Fourier transform, we then have the form of field within the waveguide as a function of $x$ and $y$,
\begin{multline}
    \phi(x,y)=-\int_{-\infty}^{\infty}\frac{{\rm d}k}{2\pi}\frac{2Kb\sin(Kd)\sin(ka)}{k\sqrt{K^2-k^2}\sin(\sqrt{K^2-k^2})}\cos(\sqrt{K^2-k^2}(1-y)){\rm e}^{{\rm i}kx}\\
 +\sum_{n=0}^{\lfloor K/\pi \rfloor}\bigg[c_n^{(+)}{\rm e}^{{\rm i}\sqrt{K^2-n^2\pi^2}x}+c_n^{(-)}{\rm e}^{-{\rm i}\sqrt{K^2-n^2\pi^2}x}\bigg]\cos(n\pi y),
\end{multline}
which is sum of two terms: the first can be interpreted as a diffraction integral, integrating over all angles of propagation away from the cavity opening, including evanescent contributions where $|K|>|k|$.  The second term is simply a sum over waveguide modes, the same solutions one would obtain if there was no connecting cavity.  As expected, in the limit of vanishing cavity depth, $d\to0$, the first term disappears from the above expression, leaving only the waveguide modes.  From this point on, for simplicity we assume that $K < \pi$ such that only the fundamental mode is supported within the waveguide.  This reduces the sum in the above expression to a single term,
\begin{multline}
\label{simplecavity-sol}
\phi = - \frac{Kab \sin(Kd)}{\pi} \lim_{\eta\to 0^+} \int_{-\infty}^\infty \D k\,\frac{\sin (ka)}{ka} \frac{\cos\left(\sqrt{(K+\I\eta)^2 - k^2}(1-y)\right)}{\sqrt{(K+\I\eta)^2 - k^2} \sin \left(\sqrt{(K+\I\eta)^2 - k^2}\right)} \E^{\I kx} + c_+ \E^{\I K x} + c_- \E^{-\I Kx},
\end{multline}
where $c_{\pm}=c_{0}^{(\pm)}$.  Note that we have evaluated the above integral as the limit $\eta\to0$---equivalent to the limit of vanishing dissipation in the air---which is required to pick out the correct (causal) branch of the square root within the integral.   As stated above, the constants $c_\pm$ can be chosen freely, but $b$ must be found from the remaining boundary conditions on the continuity of the velocity potential.  To perform this calculation in full we should do a full expansion of the cavity field as a series of all possible eigenmodes, matching the waveguide field above the cavity term--by--term.  In the approximate theory used here, where we include only a single cavity mode, we find the complex amplitude $b$ through matching the \emph{averaged} value of the velocity potential $\phi$ at the cavity entrance, which is equivalent to the inner product of the field with the spatially uniform, zeroth order mode of the cavity.

Setting $y = 0$ in both the solution (\ref{simplecavity-sol}) and the form of the field within the cavity, we integrate over the cavity opening and equate the two expressions, leaving the following form of the cavity field amplitude
\begin{equation}
    b=\frac{\sin(Ka)}{Ka}\frac{c_+ + c_-}{\cos(Kd)+ M Ka \sin(Kd)},
\end{equation}
where we have defined the integral,
\begin{equation}
    M=\frac{1}{\pi}\lim_{\eta\to 0^+} \int_{-\infty}^\infty \D k\,\left(\frac{\sin ka}{ka}\right)^2 \frac{\cot\left(\sqrt{(K+\I\eta)^2 - k^2}\right)}{\sqrt{(K+\I\eta)^2 - k^2}}\label{eq:M-def}
\end{equation}
(see Appendix \ref{app:B} for some further analysis of this expression).  This completes the solution (\ref{simplecavity-sol}).  Armed with the general solution for waves in a time-independent cavity situated as a parallel-branch to an acoustic waveguide, we now write a TMM for such a geometry, that permits the extension to an infinitely periodic crystal.

In \eqref{simplecavity-sol}, the constants $c_\pm$ are connected to the far-field, away from the cavity entrance (i. e. $|x| \gg a$).  To evaluate the field for large positive $x$, we note that the integral in (\ref{simplecavity-sol}) can be written as a contour integral, and the contour can be closed in the upper half plane.  Within this contour there is a simple pole close to the real line at $k=K+{\rm i}\eta$ plus additional poles along the imaginary axis, representing the other supported waveguide modes.  These additional poles yield a contribution to the field that exponentially decays away from the cavity opening (see Appendix~\ref{app:B}, Eq. \eqref{solution-integral-eval} for further details).  Neglecting this exponentially decaying contribution, we are left with
\begin{align}
    \phi(|x|\gg a) &= \frac{{\rm i}b}{K} \sin(Kd)\sin(Ka) \E^{\I K|x|}+ c_+ \E^{\I K x} + c_- \E^{-\I Kx}\nonumber\\
    &=-Q(c_++c_-)\E^{\I K|x|}+ c_+ \E^{\I K x} + c_- \E^{-\I Kx},\label{eq:far_left_right}
\end{align}
which we have simplified through introducing the abbreviated notation,
\begin{equation}
    Q =-{\rm i}a\left(\frac{\sin(Ka)}{K a}\right)^2\frac{1}{\cot(Kd)+ M Ka}.\label{eq:Q-def}
\end{equation}
\subsubsection{The transfer matrix}
We now write Eq. (\ref{eq:far_left_right}) in matrix form, using the notation $\phi_{R}^{(\pm)}$ for the amplitude of the left $(-)$ and right $(+)$ going parts of the field on the right of the cavity $x\gg a$, and the similar notation, $\phi_{L}^{(\pm)}$, for the field on the left.  The matrices relating these amplitudes to $c_\pm$ are,
\begin{align}
    \left(\begin{matrix}
    \phi_R^{(+)}\\\phi_R^{(-)}
    \end{matrix}\right)&= \left(\begin{matrix} 1-Q&-Q\\0& 1 \end{matrix}\right) \left(\begin{matrix} c_+\\c_-\end{matrix}\right);\nonumber \\[5pt]
    \left(\begin{matrix} \phi_L^{(+)}\\\phi_{L}^{(-)}\end{matrix}\right) &= \left(\begin{matrix}1 & 0\\-Q & 1-Q \end{matrix}\right)\left(\begin{matrix} c_+\\c_-\end{matrix}\right).\label{eq:T-matrix}
\end{align}
Combining these two equations we can eliminate the $c_\pm$ amplitudes and obtain the transfer matrix,
\begin{align} 
    \boldsymbol{T}_\text{cavity} &= \left(\begin{matrix} 1-Q&-Q \\0& 1\end{matrix}\right) \left(\begin{matrix} 1 & 0\\-Q & 1-Q\end{matrix}\right)^{-1}\nonumber\\
    &= \frac{1}{1-Q}\left(\begin{matrix}1-2Q&-Q\\Q&1\end{matrix}\right)\label{eq:T-cavity}
\end{align}
where we have chosen the convention that the transfer matrix acts on the vector of field amplitudes on the left of the cavity, yielding the amplitudes on the right. 
We note that $\det[T_\text{cavity}] = 1$, implying that the time-independent problem is always reciprocal (we show this in Appendix \ref{append:recip}, a result that carries over to the case of a single time-varying cavity). 

\subsubsection{Scattering matrix}
Having found the transfer matrix, the reflection and transmission coefficients can be found through imposing that the field amplitude vectors take the form $(\phi_{L}^{(-)},\phi_{L}^{(+)})=(r,1)$, and $(\phi_{R}^{(-)},\phi_{R}^{(+)})=(0,t)$, giving
\begin{align}
	r &= -Q \nonumber\\
	t &= 1-Q.\label{eq:r-t}
\end{align}
which are the components of the scattering matrix $S_{\rm cavity}$, relating input and output amplitudes
\begin{equation}
\begin{matrix}
    \left(\begin{matrix}\phi_{R}^{(+)}\\\phi_{L}^{(-)}\end{matrix}\right)=\left(\begin{matrix}S_{11}&S_{12}\\S_{21}&S_{22}\end{matrix}\right)\left(\begin{matrix}\phi_{L}^{(+)}\\\phi_{R}^{(-)}\end{matrix}\right)=\left(\begin{matrix}t&r\\r&t\end{matrix}\right)\left(\begin{matrix}\phi_{L}^{(+)}\\\phi_{R}^{(-)}\end{matrix}\right).
\end{matrix}
\end{equation}
For a lossless system, the scattering matrix must be unitary $S^{\dagger}S=\mathbbm{1}$, i.e.
\begin{equation}
\left(\begin{matrix}t^{\star}&r^{\star}\\r^{\star}&t^{\star}\end{matrix}\right)\left(\begin{matrix}t&r\\r&t\end{matrix}\right)=\left(\begin{matrix}|r|^2+|t|^2&t^{\star}r+r^{\star}t\\r^{\star}t+t^{\star}r&|r|^2+|t|^2\end{matrix}\right)=\left(\begin{matrix}1&0\\0&1\end{matrix}\right)
\end{equation}
From the form of the reflection and transmission coefficients (\ref{eq:r-t}), it is neither obvious that the sum of the squares of the reflection and transmission coefficients is unity, $|r|^2 + |t|^2 = 1$, nor that the real part of $r^{\star}t$ is zero.  However, the scattering matrix is unitary.  To prove this, first consider that for $|r|^2+|t|^2=1$, the above form of $r$ and $t$ requires that $|Q|^2={\rm Re}[Q]$, and that from its definition (\ref{eq:Q-def}) the real part of $Q$ is entirely due to the imaginary part of the overlap integral $M$, defined in (\ref{eq:M-def}),
\begin{equation}
    {\rm Re}[Q]=-a\left(\frac{\sin(Ka)}{Ka}\right)^2\frac{{\rm Im}[M]Ka}{|\cot(Kd)+MKa|^2}=-{\rm Im}[M]K\left(\frac{\sin(Ka)}{Ka}\right)^{-2}|Q|^2.\label{eq:Re-Q}
\end{equation}
In turn, the imaginary part of $M$ arises from the residues of the poles within the integrand of (\ref{eq:M-def}), at $k=\pm(K+{\rm i}\eta)$, which after taking the limit $\eta\to0$ is,
\begin{equation}
    {\rm Im}[M]=-\frac{1}{K}\left(\frac{\sin(Ka)}{Ka}\right)^2.\label{eq:Im-M}
\end{equation}
Combining Eqns. (\ref{eq:Im-M}) and (\ref{eq:Re-Q}) we thus prove $|Q|^2={\rm Re}[Q]$ and therefore that $|r|^2+|t|^2=1$.  Similarly $r^{\star}t+t^{\star}r=-2{\rm Re}[Q]+2|Q|^2=0$, and we thus also prove the second condition required for the scattering matrix of the resonator to be unitary.

\subsubsection{A periodic array of resonators\label{sec:periodic-array}}
\begin{figure}
    \centering
    \includegraphics[width=0.8\linewidth]{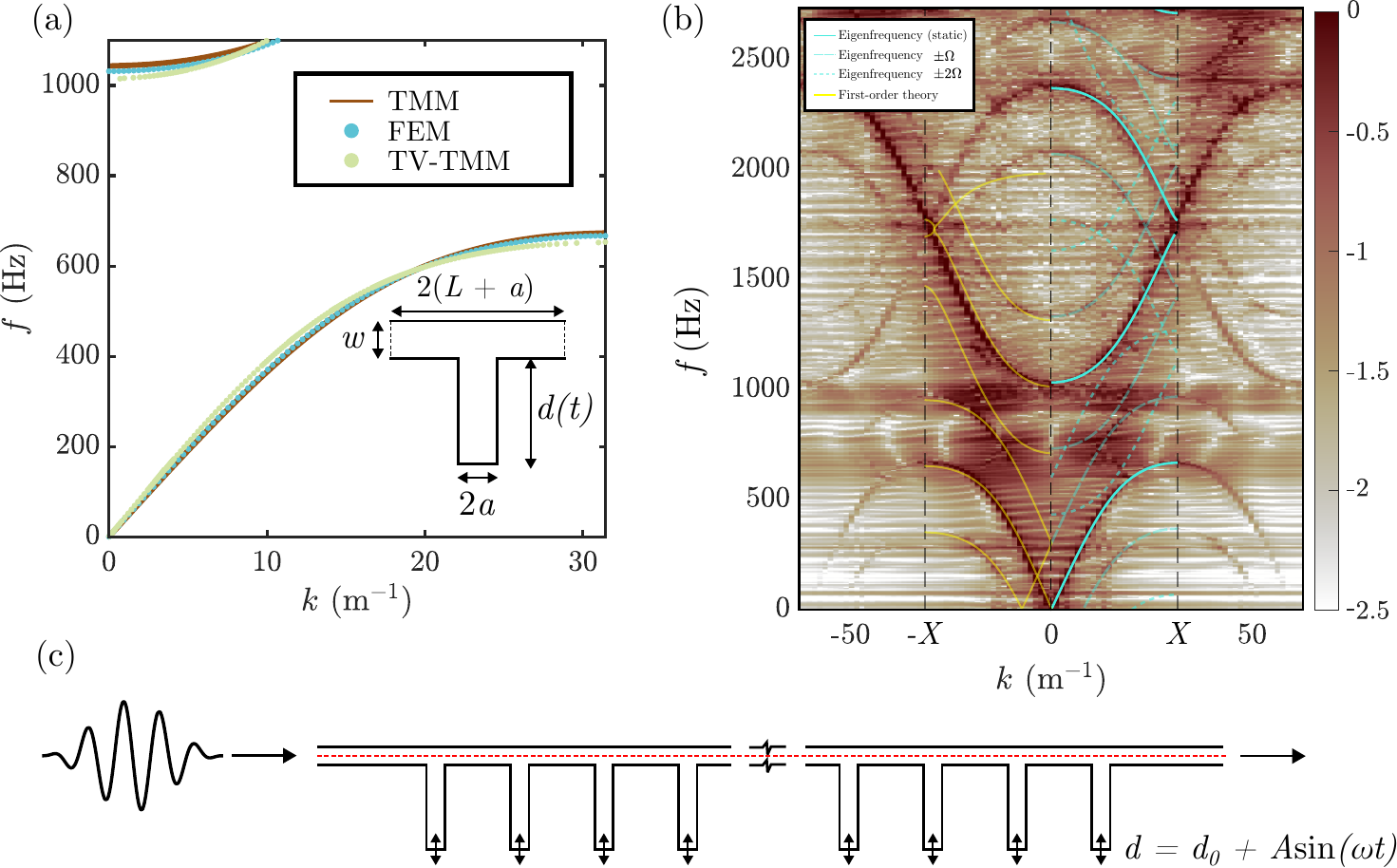}
    \caption{(a) Comparison of dispersion curves for `conventional' TMM, FEM, and time-varied TMM (TV-TMM). Inset shows unit cell that, in general, has time dependent depth, $d(t)$. Here $d(t) = d_0 = 10$ cm, $w = 2a = 2$ cm and the periodicity $2(L+a) = 10$ cm. Floquet-Bloch conditions applied at dashed lines. 
    (b) Comparison of first-order theory to FE simulations: both frequency domain (static) solutions are included, shifted by $\pm\Omega$ and $\pm 2\Omega$ (blue lines), and the Fourier spectra from a time domain simulation (colourmap). (c) Portion of the geometry in time domain simulation: a pulse is incident on a crystal comprised of 40 unit cells, whose depth changes as a function of time (same phase between all cells). A spatio-temporal Fourier transform is performed on the pressure field along the dashed red line to obtain the Fourier spectra. In all cases $w = 2a =20$ mm, $2L = 100$ mm, $d_0 = 100$ mm, $\omega = 600\pi$ rad, $A = 10$ mm. The central frequency of the pulse $\omega_0 = 1000\pi$ rad.
    }
    \label{fig:fem}
\end{figure}

Having found the transfer matrix, we can it to evaluate the dispersion relation for the Bloch wavevector, for an infinite periodic array of cavities with separation $2(L+a)$ (see Fig. \ref{fig:fem}).  To do this we enforce the Floquet-Bloch condition~\cite{jimenez2021transfer}, which corresponds to writing the two eigenvalues of of the unit cell transfer matrix as ${\rm e}^{2{\rm i}K_{\rm Bloch}(L+a)}$.  Note that the unit cell transfer matrix is not equal to the cavity transfer matrix (\ref{eq:T-cavity}), but includes extra factors due to propagation within the waveguide, $\boldsymbol{T}={\rm e}^{{\rm i}K(L+a)\boldsymbol{\sigma}_z}\boldsymbol{T}_{\rm cavity}{\rm e}^{{\rm i}K(L+a)\boldsymbol{\sigma}_z}$, where $\boldsymbol{\sigma}_{z}={\rm diag}[1,-1]$.  By definition the trace of the transfer matrix equals $2\cos(2K_{\rm Bloch}(L+a))$ which, after including the corresponding phase factors in our result (\ref{eq:T-cavity}), yields the dispersion relation
\begin{align}
    \cos(2K_{\rm Bloch}(L+a))&=\frac{1}{2}{\rm Tr}[\boldsymbol{T}]=\frac{1}{2}\left[\frac{{\rm e}^{2{\rm i}K(L+a)}}{1-Q}+\frac{(1-2Q){\rm e}^{-2{\rm i}K(L+a)}}{1-Q}\right]\nonumber\\
    &=\cos\left(2K(L+a)\right) - \frac{\I Q}{1-Q} \sin\left(2K(L+a)\right)\label{eq:disp}.
\end{align}
Noting that, our earlier result (\ref{eq:Im-M}),
\begin{equation}
\frac{Q}{1-Q}=-\frac{{\rm i}a\left(\frac{\sin(Ka)}{Ka}\right)^2}{\cot(Kd)+{\rm Re}[M]Ka+{\rm i}a\left[{\rm Im}[M]K+\left(\frac{\sin(Ka)}{Ka}\right)^{2}\right]}=-\frac{{\rm i}a\left(\frac{\sin(Ka)}{Ka}\right)^2}{\cot(Kd)+{\rm Re}[M]Ka}\label{eq:Q/1-Q}
\end{equation}
implies that $Q/(1-Q)$ is an entirely imaginary quantity, ensuring that the right hand side of the dispersion relation (\ref{eq:disp}) is always real, as expected in the absence of dissipation.  Moreover, the final expression on the right of (\ref{eq:Q/1-Q}) indicates that when $\cot(Kd)\sim-{\rm Re}[M]Ka$, the dispersion relation (\ref{eq:disp}) will exhibit a frequency gap in the dispersion relation, whenever $\sin(2K(L+a))$ is not too close to zero. 

\subsection{Scattering from an array of time-varying cavities}
\label{subsec:b}
We now present a similar analysis but allow the cavity depth to vary with time, showing that a similarly simple method can be devised.  Consider an acoustic wave propagating in the same geometry, but where now each cavity length oscillates harmonically at frequency $\Omega$.  In this section we take the cavity array to vibrate in unison with amplitude $\alpha$ and initial phase $\phi_0$. The interaction of an incoming wave with a single one of these time-varying cavities will generate subsequent waves at new frequencies (similar to a Doppler vibrometer).  We keep track of these newly generated frequency components by increasing the size of the vector of field components in our transfer matrix formalism (\ref{eq:T-matrix}).  The vector will now have $2N$ components, with $N$ of these representing the different frequencies of right--going waves, and $N$ representing the left--going waves.  Given the harmonic modulation of the cavity depth, the Bloch--Floquet theorem guarantees that the scattered frequencies from an harmonic input of frequency $\omega$ will have frequencies $\omega+n\Omega$, where $n$ is integer.
\subsubsection{Frequency conversion in a modulated cavity}
The boundary condition at the bottom of the modulated cavity $y=d(t)$ (see Fig. \ref{fig:fem}) is that the fluid velocity must equal the velocity of the bottom surface, $\dot{d}(t)$
\begin{equation}
    \left.\frac{\partial\phi}{\partial y}\right|_{y=d(t)}=\dot{d}(t)\label{eq:bc}.
\end{equation}
To fulfill the inhomogeneous boundary condition (\ref{eq:bc}), the field can broken up into two parts.  The inhomogenous solution, which satisfies (\ref{eq:bc}) with $\dot{d}$ on the right hand side, represents the sound emitted due to the modulation of the cavity depth.  Meanwhile, the homogeneous solution, with zero on the right hand, corresponds to the sound emitted from the cavity due to the incident wave and scales with the incident amplitude.  We now find the field that satisfies this boundary condition and argue that, provided the incident amplitude is large enough, we can neglect the additional sound generated from the modulation of the cavity depth.

We take the field within the cavity to be some incident field of frequency $\omega_0$ and amplitude $B_0$, plus a reflected field, written as a Fourier integral with Fourier amplitudes $\widehat{\phi}_r$,
\begin{equation}
    \phi(y,t)=B_0{\rm e}^{-\frac{{\rm i}\omega_0}{c}(y+d_0+ct)}+\int_{-\infty}^{\infty}\frac{{\rm d}\omega}{2\pi}\widehat \phi_\text{r}(\omega){\rm e}^{\frac{{\rm i}\omega}{c}(y-ct)}\label{eq:ansatz}.
\end{equation}
Inserting (\ref{eq:ansatz}) into the boundary condition (\ref{eq:bc}) we find the integral equation determining the unknown amplitudes $\widehat{\phi}_r$,
\begin{equation}
-\frac{{\rm i}\omega_0}{c}B_0{\rm e}^{-\frac{{\rm i}\omega_0}{c}(d_0+d(t)+ct)}+\int_{-\infty}^{\infty}\frac{{\rm d}\omega}{2\pi}\frac{{\rm i}\omega}{c}\widehat{\phi}_\text{r}(\omega){\rm e}^{\frac{{\rm i}\omega}{c}(d(t)-ct)}=\dot{d}(t).
\end{equation}
To invert this equation for $\widehat{\phi}_r$ we introduce the ``distorted time'' $\tau=t-d(t)/c$, noting that this is a monotonically increasing function of $t$ provided the velocity of the lower surface of the cavity never exceeds the speed of sound.  With this assumption there exists an inverse function $t(\tau)$ enabling us to perform an inverse Fourier transform
\begin{equation}
\widehat{\phi}_\text{r}(\omega)=\int_{-\infty}^{\infty}{\rm d}\tau\left[-\frac{{\rm i}c\,\dot{d}(t(\tau))}{\omega}+\frac{\omega_0}{\omega}B_0{\rm e}^{{\rm i}\omega_0(\tau-2t(\tau)-d_0/c)}\right]{\rm e}^{{\rm i}\omega\tau}.\label{eq:general-result}
\end{equation}
This is the general equation determining the reflected Fourier components from a time modulated interface, although it is often difficult to find the inverse function $t(\tau)$.  To make progress we assume that the incident velocity potential amplitude $B_0$ is large, $|B_0|\gg |c\dot{d}/\omega_0|$, such that we can neglect the sound solely produced by the time modulation of the cavity depth.  We also assume that the cavity depth modulation takes the previously stated harmonic form $d(t)=-d_0+\alpha\cos(\Omega t+\theta_0)$ such that
\begin{equation}
    \tau=t+\frac{d_0}{c} -\frac{\alpha}{c}\cos(\Omega t+\theta_0).\label{eq:tau-def}
\end{equation}
If $\alpha$ is small, the inverse function $t(\tau)$ can be approximated to leading order in $\alpha$ through substituting $t=\tau-d_0/c$ into the cosine within Eq. (\ref{eq:tau-def}).  Substituting this expression for $t(\tau)$ into our general result (\ref{eq:general-result}) and applying the generating function for Bessel functions~\cite{NIST:DLMF},
\begin{align}
\widehat{\phi}_\text{r}(\omega)&=\frac{\omega_0}{\omega}B_0{\rm e}^{\frac{{\rm i}\omega_0}{c}d_0}\int_{-\infty}^{\infty}{\rm d}\tau{\rm e}^{-2{\rm i}\omega_0\frac{\alpha}{c}\cos(\Omega(\tau-\frac{d_0}{c})+\theta_0)}{\rm e}^{{\rm i}\left(\omega-\omega_0\right)\tau}\nonumber\\
&=\frac{\omega_0}{\omega}B_0{\rm e}^{\frac{{\rm i}\omega_0}{c}d_0}\sum_{n=-\infty}^{\infty}2\pi J_{n}\left(-\frac{2\omega_0\alpha}{c}\right){\rm i}^{n}{\rm e}^{{\rm i}n\left(-\Omega\frac{d_0}{c}+\theta_0\right)}\delta(\omega-\omega_0+n\Omega).
\end{align}
which shows that the cavity modulation produces out--going waves at the set of frequencies $\omega_0+n\Omega$ (as anticipated from the Floquet--Bloch theorem), each weighted in proportion to a Bessel function of the first kind, of order $n$.  As is known from the asymptotics of Bessel functions~\cite{NIST:DLMF}, when the order $n$ increases to a value greater than the modulus of the argument $2\omega_0\alpha/c$, the Bessel functions rapidly decay with $n$.  Taking $2\omega_0\alpha/c=4\pi\alpha/\lambda\ll1$, we can truncate the series at the first harmonic $n=\pm 1$ and use the small argument approximations to the Bessel functions, $J_0(x)\sim1$ and $J_{\pm1}(x)\sim\pm x/2$, leaving
\begin{multline}
\widehat{\phi}_\text{r}(\omega)\sim2\pi\frac{\omega_0}{\omega}B_0{\rm e}^{\frac{{\rm i}\omega_0}{c}d_0}\bigg[\delta(\omega-\omega_0)
-\frac{{\rm i}\omega_0\alpha}{c}\left({\rm e}^{{\rm i}\left(\theta_0-\frac{\Omega d_0}{c}\right)}\delta(\omega-\omega_0+\Omega)+{\rm e}^{-{\rm i}\left(\theta_0-\frac{\Omega d_0}{c}\right)}\delta(\omega-\omega_0-\Omega)\right)\bigg]\label{eq:phi_w}.
\end{multline}
The aforementioned condition on the modulation distance $\alpha$ seems strict, but we shall show in the following FE simulations that the assumption gives good results even in the case where $\alpha$ is comparable to $w$.  Performing the inverse Fourier transform of (\ref{eq:phi_w}) we obtain the leading order expression for the incident and reflected fields in the modulated cavity,
\begin{multline}
    \phi(y,t)=B_0{\rm e}^{-{\rm i}K_0(y+d_0+ct)}+B_0\bigg[{\rm e}^{{\rm i}K_0(y+d_0-ct)}
    -{\rm i}\alpha\left(\frac{K_0^2}{K_-}{\rm e}^{{\rm i}\theta_0}{\rm e}^{{\rm i}K_{-}(y+d_0-ct)}+\frac{K_0^2}{K_+}{\rm e}^{-{\rm i}\theta_0}{\rm e}^{{\rm i}K_+(y+d_0-ct)}\right)\bigg].\label{eq:modulated-cavity}
\end{multline}
where $K_0 = \omega_0 / c$ and $K_\pm = \frac{\omega \pm \Omega}{c}$.  To complete the analysis, we now match the multi--frequency cavity field (\ref{eq:modulated-cavity}) to the field propagating in the main body of the wave--guide.

\subsubsection{Matching to the wave--guide mode:} \label{subsec:matching}
Setting $y=0$ in Eq. (\ref{eq:modulated-cavity}), we find the field and its normal derivative at the cavity opening in the wave--guide.  Carrying out the same analysis as in the previous section for incident waves of amplitudes $B_\pm$ and frequencies $\omega_0\pm\Omega$, we can form all of our results into a matrix relating incident amplitudes, and the three frequency components of the field at the cavity opening, $\phi_0$, $\phi_+$, and $\phi_-$.  This matrix relationship is
\begin{equation}
    \boldsymbol{\phi}_{c}=\left(\begin{matrix}\phi_-\\\phi_0\\\phi_+\end{matrix}\right)=2\boldsymbol{B}\cdot\left(\begin{matrix}B_{-}\\B_0\\B_+\end{matrix}\right),\label{eq:bc1}
\end{equation}
where the matrix $\boldsymbol{B}$ is given by
\begin{equation}\label{denote-B}
	\boldsymbol{B}= 
    \left(\begin{matrix}\cos(K_-d_0)&-\frac{{\rm i}\alpha}{2}\frac{K_0^2}{K_-}{\rm e}^{{\rm i}\theta_0}{\rm e}^{{\rm i}K_-d_0}&0\\
    -\frac{{\rm i}\alpha}{2}\frac{K_-^2}{K_0}{\rm e}^{-{\rm i}\theta_0}{\rm e}^{{\rm i}K_0 d_0}&\cos(K_0 d_0)&-\frac{{\rm i}\alpha}{2}\frac{K_+^2}{K_0}{\rm e}^{{\rm i}\theta_0}{\rm e}^{{\rm i}K_0 d_0}\\0&-\frac{{\rm i}\alpha}{2}\frac{K_0^2}{K_+}{\rm e}^{-{\rm i}\theta_0}{\rm e}^{{\rm i}K_+d_0}&\cos(K_+d_0)\end{matrix}\right).
\end{equation}
Similarly, for the normal derivative of the field at the cavity opening we have a matrix relation between frequency components of the derivative $\phi'_-$, $\phi'_0$, and $\phi'_+$, and the same set of incident amplitudes,
\begin{equation}
    \boldsymbol{\phi}'_c=\left(\begin{matrix}\phi_-'\\\phi_0'\\\phi_+'\end{matrix}\right)=2\boldsymbol{B}'\cdot\left(\begin{matrix}B_{-}\\B_0\\B_+\end{matrix}\right)=2\frac{\partial\boldsymbol{B}}{\partial d_0}\cdot\left(\begin{matrix}B_{-}\\B_0\\B_+\end{matrix}\right).\label{eq:bc2}
\end{equation}
The procedure is now a generalization of that described in Sec. \ref{sec:time-independent}, where we eliminate the unknowns $B_-$, $B_0$ and $B_+$.  We can do this through combining Eqns. (\ref{eq:bc1}--\ref{eq:bc2}), finding an effective impedance boundary condition at the cavity opening
\begin{equation}
    \boldsymbol{\phi}'_c=\frac{\partial\boldsymbol{B}}{\partial d_0}\cdot\boldsymbol{B}^{-1}\cdot\boldsymbol{\phi}_c
\label{eq:impedance}\end{equation}
Expanding the field in the waveguide as in Sec.~\ref{sec:time-independent}, using the above vector description to account for the set of frequencies involved,
\begin{multline}
    \boldsymbol{\phi}(x,y)=\lim_{\eta\to0}\int_{-\infty}^{\infty}\frac{{\rm d}k}{2\pi}{\rm e}^{{\rm i}kx}\frac{\sin(ka)}{k}\frac{\cos(\sqrt{(\boldsymbol{K}+{\rm i}\eta\mathbbm{1})^2-k^2\mathbbm{1}}(1-y))}{\sqrt{(\boldsymbol{K}+{\rm i}\eta\mathbbm{1})^2-k^2\mathbbm{1}}\sin(\sqrt{(\boldsymbol{K}+{\rm i}\eta\mathbbm{1})^2-k^2\mathbbm{1}})}\cdot\boldsymbol{\phi}_{c}'\\
 +{\rm e}^{{\rm i}\boldsymbol{K}x}\cdot\boldsymbol{c}^{(+)}+{\rm e}^{-{\rm i}\boldsymbol{K}x}\cdot\boldsymbol{c}^{(-)},\label{eq:phi-vec}
\end{multline}
where $\boldsymbol{K}={\rm diag}[K_-,K_0,K_+]$, and we have imposed the boundary conditions of vanishing normal derivative on the upper surface ($y=1$) and of vanishing normal derivative except at the cavity aperture on the lower surface ($y=0$).  As all matrices are diagonal in Eq. (\ref{eq:phi-vec}) there is no ambiguity in their order and we have therefore used the same notation for division etc. as used for ordinary numbers.

To determine the unknown vector $\boldsymbol{\phi}'_c$ (or equivalently $\boldsymbol{\phi}_c$) in Eq. (\ref{eq:phi-vec}) we impose the generalized impedance boundary condition (\ref{eq:impedance}) on the value of the velocity potential at the opening of the cavity, again averaging through integrating over the aperture.  This gives the following condition
\begin{equation}
    \boldsymbol{\phi}_c=a\boldsymbol{M}\cdot\frac{\partial\boldsymbol{B}}{\partial d_0}\cdot\boldsymbol{B}^{-1}\cdot\boldsymbol{\phi}_c\\
 +\frac{2\sin(\boldsymbol{K}a)}{\boldsymbol{K}a}\cdot\left(\boldsymbol{c}^{(+)}+\boldsymbol{c}^{(-)}\right)\label{eq:phi-c}
\end{equation}
where we have defined the multi--frequency generalization of the overlap integral defined for time--independent cavities (\ref{eq:M-def}),
\begin{equation}
    \boldsymbol{M}=\frac{1}{\pi}\lim_{\eta\to0}\int_{-\infty}^{\infty}{\rm d}k\left(\frac{\sin(ka)}{ka}\right)^2\frac{\cot(\sqrt{(\boldsymbol{K}+{\rm i}\eta\mathbbm{1})^2-k^2\mathbbm{1}})}{\sqrt{(\boldsymbol{K}+{\rm i}\eta\mathbbm{1})^2-k^2\mathbbm{1}}}. \label{eq:M-vector}
\end{equation}
The solution for the unknown $\boldsymbol{\phi}_c$ appearing in Eq. (\ref{eq:phi-c}) is found through a simple re--arrangement and matrix inversion
\begin{equation}
    \boldsymbol{\phi}_c=\left[\mathbbm{1}-a\boldsymbol{M}\cdot\frac{\partial\boldsymbol{B}}{\partial d_0}\cdot\boldsymbol{B}^{-1}\right]^{-1}\cdot\frac{2\sin(\boldsymbol{K}a)}{\boldsymbol{K}a}\cdot\left(\boldsymbol{c}^{(+)}+\boldsymbol{c}^{(-)}\right)\label{eq:phi-c}
\end{equation}
which completes the solution of the problem of matching the waveguide field to the cavity.  Note that, through Eq. (\ref{eq:impedance}) we can also use Eq. (\ref{eq:phi-c}) to find the components of the normal velocity at the cavity aperture, $\boldsymbol{\phi}'_c$.

\subsubsection{Transfer matrix formalism}

We shall now develop the TM in the waveguide.  We first substitute our solution (\ref{eq:phi-c}) for the unknown velocity and velocity potentials at the entrance to the cavity into our general expression for the field in the waveguide (\ref{eq:phi-vec}).  Despite first appearances, the integrand in Eq. (\ref{eq:phi-vec}) is actually free of branch cuts and when $|x|\gg a$ we can close the contour in one half of the $k$ plane (upper for $x>0$, lower for $x<0$), leaving the form of the velocity potential in the wave--guide either side of the cavity aperture,
\begin{equation}
    \boldsymbol{\phi}(x,y)=-{\rm e}^{{\rm i}\boldsymbol{K}|x|}\cdot\boldsymbol{Q}\cdot\left(\boldsymbol{c}^{(+)}+\boldsymbol{c}^{(-)}\right)
    +{\rm e}^{{\rm i}\boldsymbol{K}x}\cdot\boldsymbol{c}^{(+)}+{\rm e}^{-{\rm i}\boldsymbol{K}x}\cdot\boldsymbol{c}^{(-)},\label{eq:vec-phi-lim}
\end{equation}
where we have defined the matrix quantity $\boldsymbol{Q}$, analogous to our previous definition (\ref{eq:Q-def}),
\begin{equation}
    \boldsymbol{Q}={\rm i}a\frac{1}{\boldsymbol{K}}\cdot\frac{\sin(\boldsymbol{K}a)}{\boldsymbol{K}a}\cdot\frac{\partial\boldsymbol{B}}{\partial d_0}\cdot\boldsymbol{B}^{-1}\cdot\left[\mathbbm{1}-a\boldsymbol{M}\cdot\frac{\partial\boldsymbol{B}}{\partial d_0}\cdot\boldsymbol{B}^{-1}\right]^{-1}\cdot\frac{\sin(\boldsymbol{K}a)}{\boldsymbol{K}a}\label{eq:M-Q-def}
\end{equation}
The transfer matrix formalism now follows, once we have found the relationship between the left and right going parts of the field (each now consisting of three different frequencies), either side of the cavity entrance.  Using an extension of the notation used earlier in Eq. (\ref{eq:T-matrix}) we can use Eq. (\ref{eq:vec-phi-lim}) to separate out the left and right--going parts of the field either side of the cavity,
\begin{equation}
\left(\begin{matrix}\boldsymbol{\phi}^{(+)}_{R}\\\boldsymbol{\phi}^{(-)}_{R}\end{matrix}\right)=\left(\begin{matrix}\mathbbm{1}-\boldsymbol{Q}&-\boldsymbol{Q}\\\boldsymbol{0}&\mathbbm{1}\end{matrix}\right)\left(\begin{matrix}\boldsymbol{c}^{(+)}\\\boldsymbol{c}^{(-)}\end{matrix}\right)\label{eq:subT1}
\end{equation}
and
\begin{equation}
\left(\begin{matrix}\boldsymbol{\phi}^{(+)}_{L}\\\boldsymbol{\phi}^{(-)}_{L}\end{matrix}\right)=\left(\begin{matrix}\mathbbm{1}&\boldsymbol{0}\\-\boldsymbol{Q}&\mathbbm{1}-\boldsymbol{Q}\end{matrix}\right)\left(\begin{matrix}\boldsymbol{c}^{(+)}\\\boldsymbol{c}^{(-)}\end{matrix}\right).\label{eq:subT2}
\end{equation}
Eliminating the vector $(\boldsymbol{c}^{(+)},\boldsymbol{c}^{(-)})$, we then form the transfer matrix, as before:
\begin{align}
    \boldsymbol{T}_{\rm cavity}&=\left(\begin{matrix}\mathbbm{1}-\boldsymbol{Q}&-\boldsymbol{Q}\\\boldsymbol{0}&\mathbbm{1}\end{matrix}\right)\left(\begin{matrix}\mathbbm{1}&\boldsymbol{0}\\-\boldsymbol{Q}&\mathbbm{1}-\boldsymbol{Q}\end{matrix}\right)^{-1}\nonumber\\
    &=\left(\begin{matrix}2\mathbbm{1}-(\mathbbm{1}-\boldsymbol{Q})^{-1}&\mathbbm{1}-(\mathbbm{1}-\boldsymbol{Q})^{-1}\\
    (\mathbbm{1}-\boldsymbol{Q})^{-1}-\mathbbm{1}&(\mathbbm{1}-\boldsymbol{Q})^{-1}\end{matrix}\right),\label{eq:T-cav-final}
\end{align}
which can be applied in sequence to find the propagation characteristics of an arbitrary sequence of modulated cavities.  We note that, although a single modulated cavity will, due to the symmetry $x\leftrightarrow-x$, exhibit reciprocity, where the transmission matrix for propagation from left to right equals that for propagation from right to left, two or more cavities with a relative phase difference in their time modulation will no longer necessarily exhibit reciprocal transmission.  This is discussed further in appendix \ref{append:recip}.

\subsubsection{Scattering matrix}
The scattering matrix is now also a $2N\times2N$ object, containing reflection and transmission \emph{matrices}, reflecting the coupling between frequencies due to the modulation of the cavity,
\begin{equation}
    \boldsymbol{S}=\left(\begin{matrix}\boldsymbol{t}&\boldsymbol{r}\\\boldsymbol{r}&\boldsymbol{t}\end{matrix}\right).
\end{equation}
As before, this matrix relates the outgoing waves to the incoming ones, as follows,
\begin{equation}
    \left(\begin{matrix}\boldsymbol{\phi}_{R}^{(+)}\\\boldsymbol{\phi}_{L}^{(-)}\end{matrix}\right)=\left(\begin{matrix}\boldsymbol{t}&\boldsymbol{r}\\\boldsymbol{r}&\boldsymbol{t}\end{matrix}\right)\left(\begin{matrix}\boldsymbol{\phi}_{L}^{(+)}\\\boldsymbol{\phi}_{R}^{(-)}\end{matrix}\right).
\end{equation}
The elements of the scattering matrix can be determined from the transfer matrix (\ref{eq:T-cav-final}) either through re--writing the vectors on the left of Eqns. (\ref{eq:subT1}--\ref{eq:subT2}) in terms of the input and output vectors, then forming the scattering rather than transfer matrix, or through applying the transfer matrix separately to field vectors with zero incoming waves from the left or right, finding the transmission and reflection matrices in the process.  Following the second procedure, we find the transmission and reflection matrices equal
\begin{align}
    \boldsymbol{r}&=-\boldsymbol{Q}\nonumber,\\[5pt]
    \boldsymbol{t}&=\mathbbm{1}-\boldsymbol{Q}.
\end{align}
Notice that, even with lossless material parameters, in the case of a time--modulated medium, $||\vv r||^2 + ||\vv t||^2 = 1$ is not a necessity; there is a source of energy inside each cavity arising from the temporal variation of the cavity length.

\section{Space-time modulated acoustic metamaterials}

We now employ the above developed time--varying transfer matrix method to obtain the dispersion relation for space-time modulated materials. We treat the simplest case: an infinite array of the cavities described above, with depths (resonant frequencies) modulated in phase, with identical amplitude. 

As in Sec. \ref{sec:time-independent}, the transfer matrix for a unit cell of the waveguide can be obtained through introducing the propagation phase either side of the cavity,
\begin{align}
    \boldsymbol{T}_{\text{Unit cell}}&=\left(\begin{matrix}{\rm e}^{{\rm i}\boldsymbol{K}(L+a)}[2\mathbbm{1}-(\mathbbm{1}-\boldsymbol{Q})^{-1}]{\rm e}^{{\rm i}\boldsymbol{K}(L+a)}&{\rm e}^{{\rm i}\boldsymbol{K}(L+a)}[\mathbbm{1}-(\mathbbm{1}-\boldsymbol{Q})^{-1}]{\rm e}^{-{\rm i}\boldsymbol{K}(L+a)}\\
    {\rm e}^{-{\rm i}\boldsymbol{K}(L+a)}[(\mathbbm{1}-\boldsymbol{Q})^{-1}-\mathbbm{1}]{\rm e}^{{\rm i}\boldsymbol{K}(L+a)}&{\rm e}^{-{\rm i}\boldsymbol{K}(L+a)}(\mathbbm{1}-\boldsymbol{Q})^{-1}{\rm e}^{-{\rm i}\boldsymbol{K}(L+a)}\end{matrix}\right)\nonumber\\
    &=\left(\begin{matrix}\boldsymbol{T}_{11}&\boldsymbol{T}_{12}\\\boldsymbol{T}_{21}&\boldsymbol{T}_{22}\end{matrix}\right).
    \label{eq:T-unit-cell}
\end{align}
The six eigenvalues of the $6\times6$ transfer matrix (\ref{eq:T-unit-cell}) represent left and right propagation for the three branches of the  dispersion relation in this time--modulated system.  In general these can be numerically determined from Eq. (\ref{eq:T-unit-cell}), but we can simplify the problem here through taking advantage of the reflectional symmetry;  inverting the coordinate system $x\to-x$ around the centre of one of the cavities simply changes the sign of the Bloch--vector and interchanges the left and right--going waves.  Meanwhile, due to the symmetry of the system, the transfer matrix remains unchanged.  This indicates that the eigenvectors of (\ref{eq:T-unit-cell}) come in pairs $\boldsymbol{v}_{\pm}$ related by (see Appendix \ref{append:recip} for a formal proof),
\begin{equation}
    \boldsymbol{v}_{-}=\boldsymbol{U}\cdot\boldsymbol{v}_{+}=\left(\begin{matrix}\boldsymbol{0}&\mathbbm{1}\\\mathbbm{1}&\boldsymbol{0}\end{matrix}\right)\boldsymbol{v}_{+},\qquad{\rm Eigenvalues:}\;\lambda_{\pm}={\rm e}^{\pm2{\rm i}K_{\rm Bloch}(L+a)}.\label{eq:transformation}
\end{equation}
Therefore if we add together the transfer matrix (\ref{eq:T-unit-cell}) and the transformed matrix $\boldsymbol{U}\cdot\boldsymbol{T}_{\rm Unit cell}\cdot\boldsymbol{U}^{\dagger}$,
\begin{align}
    \boldsymbol{T}_{\text{Unit cell}}'&=\frac{1}{2}\left[\left(\begin{matrix}\boldsymbol{T}_{11}&\boldsymbol{T}_{12}\\\boldsymbol{T}_{21}&\boldsymbol{T}_{22}\end{matrix}\right)+\left(\begin{matrix}\boldsymbol{0}&\mathbbm{1}\\\mathbbm{1}&\boldsymbol{0}\end{matrix}\right)\left(\begin{matrix}\boldsymbol{T}_{11}&\boldsymbol{T}_{12}\\\boldsymbol{T}_{21}&\boldsymbol{T}_{22}\end{matrix}\right)\left(\begin{matrix}\boldsymbol{0}&\mathbbm{1}\\\mathbbm{1}&\boldsymbol{0}\end{matrix}\right)\right]\nonumber\\
    &=\frac{1}{2}\left(\begin{matrix}\boldsymbol{T}_{11}+\boldsymbol{T}_{22}&\boldsymbol{T}_{12}+\boldsymbol{T}_{21}\\\boldsymbol{T}_{12}+\boldsymbol{T}_{21}&\boldsymbol{T}_{11}+\boldsymbol{T}_{22}\end{matrix}\right),
\end{align}
we obtain a matrix with degenerate pairs of eigenvalues, $\frac{1}{2}(\lambda_++\lambda_-)=\cos(2K_{\rm Bloch}(L+a))$, which can be reduced to block diagonal form after transformation into symmetric and anti-symmetric combinations of left and right--going waves,
\begin{align}
    \frac{1}{\sqrt{2}}\left(\begin{matrix}\mathbbm{1}&\mathbbm{1}\\\mathbbm{1}&-\mathbbm{1}\end{matrix}\right)\frac{1}{2}\left(\begin{matrix}\boldsymbol{T}_{11}+\boldsymbol{T}_{22}&\boldsymbol{T}_{12}+\boldsymbol{T}_{21}\\\boldsymbol{T}_{12}+\boldsymbol{T}_{21}&\boldsymbol{T}_{11}+\boldsymbol{T}_{22}\end{matrix}\right)\frac{1}{\sqrt{2}}\left(\begin{matrix}\mathbbm{1}&\mathbbm{1}\\\mathbbm{1}&-\mathbbm{1}\end{matrix}\right)\\
    =\frac{1}{2}\left(\begin{matrix}\boldsymbol{T}_{11}+\boldsymbol{T}_{22}+\boldsymbol{T}_{12}+\boldsymbol{T}_{21}&\boldsymbol{0}\\\boldsymbol{0}&\boldsymbol{T}_{11}+\boldsymbol{T}_{22}-\boldsymbol{T}_{12}-\boldsymbol{T}_{21}\end{matrix}\right).\label{eq:Tpp}
\end{align}
Having reduced the transfer matrix to this block diagonal form, we can now find the dispersion relation through analyzing the eigenvalues of the $3\times 3$ block matrices, as these come in degenerate pairs.  Taking the lower block of Eq.(\ref{eq:Tpp}) we obtain the dispersion relation,
\begin{multline}
    {\rm det}\left[\frac{1}{2}\left(\boldsymbol{T}_{11}+\boldsymbol{T}_{22}-\boldsymbol{T}_{12}-\boldsymbol{T}_{21}\right)\right] = 0\\
    ={\rm det}\bigg[\mathbbm{1}\cos(2K_{\rm Bloch}(L+a))-{\rm e}^{2{\rm i}\boldsymbol{K}(L+a)}
    +2{\rm i}\cos(\boldsymbol{K}(L+a))(\mathbbm{1}-\boldsymbol{Q})^{-1}\sin(\boldsymbol{K}(L+a))\bigg]\label{eq:bloch-dispersion-det}.
\end{multline}

From Eq. (\ref{eq:bloch-dispersion-det}), we see that if we turn off all cavity modulations (such that the matrix $\boldsymbol{Q}$ becomes diagonal), we recover our previous dispersion relation (\ref{eq:disp}) for the three separate frequencies, $\omega$, $\omega+\Omega$, and $\omega-\Omega$. However, here we obtain three eigenvalues; for each (central) frequency, we obtain (up to) three different waves that can propagate through an infinite array of vibrating cavities. 

Crucially, if $\alpha \ll 1$, it follows from \eqref{denote-B} that the off-diagonal entries in the matrices $B$ and $\pdv{B}{d}$ are  $O(\alpha)$. We can quickly convince ourselves that sums, products and inverses of matrices with $O(\alpha)$ off-diagonal entries are once more matrices with $O(\alpha)$ off-diagonal entries, and so from \eqref{eq:M-Q-def} we see that $Q$, and the matrix in \eqref{eq:bloch-dispersion-det}, are of such form as well.

These matrices have another interesting property: their determinant is equal to the product of the diagonal entries, plus contributions of \emph{second} and higher orders. This is true because the determinant is a sum of all possible products obtained by taking exactly one entry from each row and each column. Obviously, we cannot form such products with exactly one off-diagonal entry (because after choosing $n-1$ diagonal entries, there is only one possible choice of the last entry, and it is the remaining diagonal entry).

This leads to an important observation: \emph{adding an $O(\alpha)$ off-diagonal perturbation to a diagonal matrix whose entries are all different changes its eigenvalues only in the \emph{second} order}, because its eigenvalues are given by the equation $\det(M - \lambda) = 0$, and, according to the previous fact, $\det(M - \lambda) = (M_{11} - \lambda) (M_{22} - \lambda) \cdots (M_{nn} - \lambda) + O(\alpha^2)$. If $\alpha = 0$, then the eigenvalues are exactly equal to the diagonal entries; if it is very small, the eigenvalue $M_{kk}$ will shift by a very small amount $\delta$. Substituting into the equation, we have
\[ (M_{11} - M_{kk} - \delta) \cdots (M_{kk} - M_{kk} - \delta) \cdots (M_{nn} -M_{kk}-\delta) + O(\alpha^2) = 0, \]
and if all $M_{kk}$ are different, we get $\delta = 0$ up to the first order.

From this it immediately follows that \emph{up to the first order in $\alpha$}, turning on the vibration has no effect on the shape of the dispersion curves, because the small off-diagonal perturbation does not shift the eigenvalues. Hence the eigenvalues will only change by an $O(\alpha^2)$ contribution. Hence, with sufficiently small vibrations, the curves in the Bloch dispersion relation will look the same, apart from a shift; each resonator (when modulated in phase) acts as a secondary Doppler source of sum and difference frequencies. 

Finally, let us consider what happens if two of the dispersion curves cross. In that case, it is no longer true that the diagonal entries are all different, and we must adjust our reasoning slightly. For concreteness, we work out the case of $3\times3$ matrices. Suppose that $M_{11} = M_{22} = M$ and we would like to know the shift of this double eigenvalue. We let $\lambda = M+\delta$, and then the eigenvalue equation reduces to $\delta^2 (M_{33} - M) + O(\alpha^2) = 0$ -- $O(\alpha^2)$ terms may no longer be neglected. However, after expanding the determinant we find, up to $O(\alpha^2)$, the characteristic equation reduces to
\[ \delta^2(M_{33}-M) - (M_{33}-M) \alpha^2 M_{12}M_{21} = 0, \]
and (if $M_{33} \ne M$), we find that the double eigenvalue $M$ splits into two different eigenvalues, $M \pm \alpha \sqrt{M_{12}M_{21}}$. This is true for $3\times3$ matrices only. For larger matrices, the result would be similar (a double eigenvalue would split into two with the shift being proportional to $\alpha$) but the expression under the square root would be much more complicated.

Hence in the case of the two dispersion curves crossing, the double eigenvalue suddenly splits into two. The amount of splitting can, in principle, be obtained by expanding the matrix in \eqref{eq:bloch-dispersion-det} up to the first order in $\alpha$. We remark that the eigenvalue shifts are in general complex, and find that small band gaps can therefore open at crossings of different curves of the dispersion relation.

We focus here on showing the efficacy of this model, in predicting the first order shifts of the dispersion curves by comparison to FE simulations, shown in Fig.~\ref{fig:fem}. Using the acoustics module in COMSOL Multiphysics \cite{Comsol} we perform two simulations; firstly, a static (lossless) frequency domain simulation on a unit cell with geometrical parameters outlined in Fig.~\ref{fig:fem}, the results of this are then then simply shifted by the Doppler terms $\pm \Omega, 2\Omega$ as we predict assuming the amplitude of oscillation is small; Secondly, we perform a time domain simulation (incorporating thermo-viscous losses on the boundaries) of a waveguide with 40 unit cells of side-branching QWRs, and we neglect the motion of the boundaries as a source of acoustic radiation. A pulse centered on $\omega_0$ is injected and allowed to propagate through the crystal, with perfectly matched layers (PMLS) at either end. A spatio-temporal Fourier transform is applied to the acoustic pressure field extracted along the centre line of the waveguide, and the resulting Fourier spectra shown as the colour map in Fig.~\ref{fig:fem}. Clear matches between the `copied' dispersion curves from the static simulation can be seen on the RHS of the dispersion curve. In yellow, on the LHS, we overlay the resulting dispersion curves from the temporally-varied TMM and see that, despite the simplifying assumptions, the solution to the ensuing eigenvalue problem returns good agreement up to the second branch (and its Doppler shifts). Therefore, we assert that the TV-TMM presented here serves as a building block towards space-time modulated metamaterials that can be used to design, in principle, non-reciprocal systems (highlighted in Appendix~\ref{append:recip}) and even moving gratings \cite{horsley2024traveling}.

\section{Conclusion}

We have constructed a transfer matrix theory, applicable to a one dimensional wave-guide containing a sequence of side-branching quarter wavelength resonators, each modulated at frequency $\Omega$.  In the above formalism, we constructed the transfer matrix through an approximate mode-matching at the boundary of each cavity, incorporating temporal variations of the fundamental cavity resonant frequency by periodically varying the cavity length at frequency $\Omega$.  A more accurate treatment can be obtained through the inclusion of further cavity modes, which will modify the expression for the coupling matrix $\boldsymbol{M}$, defined in Eq. (\ref{eq:M-def}).  Similarly, the accuracy of the method can be applied to more extreme modulations of the cavity through simply increasing the number of frequencies from the three considered here, to $N$, resulting in a $2N\times2N$ transfer matrix.  Considering the special case of a periodic array of such resonators, we have validated the predictions of our theory via finite element simulations, showing excellent agreement both in the frequency and time-domain.  This theory, like the conventional TMM, will be a useful tool in the modelling of time-varying acoustic metamaterials, expecially in cases where the modulation varies inhomogeneously in both time and space, making numerical simulations intractable, or at least very time consuming.

\vskip6pt

\enlargethispage{20pt}

\section*{Acknowledgments}
S.A.R.H and G.J.C acknowledge the financial support by the EPSRC (grant no EP/Y015673/1). 
‘For the purpose of open access, the author has applied a ‘Creative Commons Attribution (CC BY) licence to any Author Accepted Manuscript version arising from this submission’.

\appendix
\counterwithin*{equation}{section}
\renewcommand\theequation{\thesection\arabic{equation}}

\section{Conventional TMM approach}
\label{ap:tmm}
We benchmark the static modal-matched TMM against the classic TMM where the acoustic state vector $\left( p, v\right)^T$ at position $x_1$ is related to the state vector at $x_2$, i.e. 
\begin{equation}
    \begin{pmatrix}
        p \\ v
    \end{pmatrix}_{x_1} = 
    \begin{pmatrix}
        T_{11} &T_{12} \\ 
        T_{21} &T_{22}
    \end{pmatrix}
    \begin{pmatrix}
        p \\ v
    \end{pmatrix}_{x_2}
\end{equation}
The cavity (a quarter wavelength resonator (QWR)) is modeled with an effective specific cavity impedance $Z_c^{\prime}$ such that $Z_c^{\prime} = iZ_{c}\cot(K\ell)/A_c$, neglecting end corrections, with $\ell = d_0$ and $A_c$ the cross sectional area of the cavity, and $Z_c$ is the characteristic impedance of the fluid in the cavity. In general $K$ is an effective complex wavenumber $K_\text{eff}(\omega)$ when visco-thermal losses are considered in the cavity (which we neglect here, as in the frequency domain FE simulation). The transfer matrix of the repeated unit cell then consists of the products of propagator matrices $T_{\text{prop}}$ and a parallel side-duct  $T_{\text{cavity}}$ \cite{jimenez2021transfer}
\begin{align}
\begin{split}
     \begin{pmatrix}
        T_{11} &T_{12} \\ 
        T_{21} &T_{22}
    \end{pmatrix} 
    = \underbrace{\begin{pmatrix}
        \cos(K(L+a)) &-iZ_0\sin(K(L+a)) \\
        \frac{-i}{Z_0}\sin(K(L+a))  &\cos(K(L+a))
    \end{pmatrix}}_{T_\text{prop}} \cdotp
    \underbrace{\begin{pmatrix}
        1 & 0 \\
        1/Z_c^{\prime} &1
    \end{pmatrix}}_{T_{\text{cavity}}} \\
    \cdotp \underbrace{\begin{pmatrix}
        \cos(K(L+a)) &-iZ_0\sin(K(L+a)) \\
        \frac{-i}{Z_0}\sin(K(L+a)) & \cos(K(L+a))
    \end{pmatrix}}_{T_\text{prop}},
    \end{split}
    \end{align}
where $Z_0$ is the impedance of the waveguide (again assuming no loss) i.e. $Z_0 = \rho_0c_0/A_0$ with $\rho_0$ the density and $c_0$ the sound speed, and $A_0$ is the cross-sectional area of the guide. Here both the cavity and waveguide are assumed to be rectangular (with dimension out of the plane 10 m, to approximate a 2D guide). Employing the Floquet-Bloch conditions then results in a convenient form of the dispersion relation, relating to the trace of the transfer matrix:
\begin{equation}
    \cos(2K(L+a)) = \frac{T_{11} + T_{22}}{2},
\end{equation}
which we show as a useful comparison in Fig 1.

\section{Useful formulas for integrals occurring in the solution}
\label{app:B}
The integrals appearing in the above solutions---i.e. the integral expression for the field (\ref{simplecavity-sol}) along with Eq. (\ref{eq:M-def}) and its matrix generalization---are often unwieldy and so here we present a set of results for solving them analytically.  Firstly we note that the integral in (\ref{simplecavity-sol}) can be split into real and imaginary parts through taking the limit $\eta\to 0$ (thus moving the pole onto the real $k$ axis),
\begin{align}
    \begin{split}
         \lim_{\eta\to 0^+} \int_{-\infty}^\infty \frac{\sin ka}{ka} \frac{\cos[\sqrt{(K+\I\eta)^2 - k^2}(1-y)]}{\sqrt{(K+\I\eta)^2 - k^2} \sin \sqrt{(K+\I\eta)^2 - k^2}} \E^{\I kx}\,\D k  \\= - \I\pi \frac{\sin Ka \cos Kx}{K^2 a} + \text{P} \int_{-\infty}^\infty \frac{\sin ka}{ka} \frac{\cos[\sqrt{K^2 - k^2}(1-y)]}{\sqrt{K^2 - k^2} \sin \sqrt{K^2 - k^2}} \E^{\I kx}\,\D k \label{eq:res-1}
    \end{split}
\end{align}
where `$\text{P}$' indicates the Cauchy principal value of the integral.  Secondly, we remark the above integrand on the second line of (\ref{eq:res-1}) is even in $k$.  Hence, it may also be written as
\[ 2\ \text{P} \int_0^\infty \frac{\sin ka}{ka} \frac{\cos[\sqrt{K^2 - k^2}(1-y)]}{\sqrt{K^2 - k^2} \sin \sqrt{K^2 - k^2}} \cos kx\,\D k. \]
We also note that, in case of $|x| > a$, the residue theorem can be applied to (\ref{eq:res-1}), transforming the integral into a summation,
\begin{multline} \label{solution-integral-eval}
	\lim_{\eta\to 0^+} \int_{-\infty}^\infty \frac{\sin ka}{ka} \frac{\cos[\sqrt{(K+\I\eta)^2 - k^2}(1-y)]}{\sqrt{(K+\I\eta)^2 - k^2} \sin \sqrt{(K+\I\eta)^2 - k^2}} \E^{\I kx}\,\D k \\
	= \pi \frac{\sin Ka}{Ka} \frac{\E^{\I K|x|}}{\I K} - \frac{2\pi}{a} \sum_{n=1}^\infty (-1)^n \frac{\sinh(a \sqrt{n^2\pi^2 - K^2}) \cos[n\pi(1-y)]}{n^2 \pi^2 - K^2} \E^{-\sqrt{n^2\pi^2 - K^2}|x|}.
\end{multline}
The integral that defines $M$ in \eqref{eq:M-def} can be evaluated by the residue theorem as well, giving the following series expansion
\begin{equation} \label{m-eval}
	M = \frac{\cot K}{Ka} - \frac{\E^{\I K a}}{K^2 a} \frac{\sin Ka}{Ka} + \frac{1}{a^2} \sum_{n=1}^\infty \frac{1 - \exp(-2a\sqrt{n^2 \pi^2 - K^2})}{(n^2 \pi^2 - K^2)^{3/2}}.
\end{equation}

\section{Directional symmetry and non-reciprocity}
\label{append:recip}
Here we outline additional uses of the TV-TMM developed. We start by showing the directional symmetry of a single time-varying cavity, before detailing the procedure to produce non-reciprocal transmission. 

We firstly apply the unitary transformation $\boldsymbol{U}$ defined in Eq. (\ref{eq:transformation}), which switches left-- and right--going waves. We note that multiplying a block matrix by this from the left will switch around the ``block rows'', and doing the same from the right will switch the ``block columns''. As a result we have $\boldsymbol{U}^2 = 1$.

As discussed in the main text, this means that the transfer matrix in \eqref{eq:T-cav-final} can be written as $\boldsymbol{T}_\text{cavity} = \boldsymbol{M}(\boldsymbol{U}\boldsymbol{M}\boldsymbol{U})^{-1} = \boldsymbol{M}\boldsymbol{U}\boldsymbol{M}^{-1}\boldsymbol{U}$, and $\boldsymbol{T}_\text{cavity}^{-1} = \boldsymbol{U}\boldsymbol{M}\boldsymbol{U}\boldsymbol{M}^{-1}$. Multiplying this by $\boldsymbol{U}$ from both sides, we get $\boldsymbol{U}\boldsymbol{T}_\text{cavity}^{-1} \boldsymbol{U} = \boldsymbol{M}\boldsymbol{U}\boldsymbol{M}^{-1} \boldsymbol{U} = \boldsymbol{T}_\text{cavity}$, and so
\[ \boldsymbol{T}_\text{cavity}^{-1} = \boldsymbol{U}\boldsymbol{T}_\text{cavity}\boldsymbol{U}. \]
Now let $\vv v$ be an eigenvector of $\boldsymbol{T}_\text{cavity}$ with an eigenvalue of $\lambda$. Then we have
\[ \boldsymbol{T}_\text{cavity} \vv v = \lambda \vv v \quad\implies\quad \boldsymbol{T}_\text{cavity}^{-1} \vv v = \lambda^{-1} \vv v \quad\implies\quad \boldsymbol{U}\boldsymbol{T}_\text{cavity} \boldsymbol{U} \vv v = \lambda^{-1} \vv v \quad\implies\quad \boldsymbol{T}_\text{cavity} \boldsymbol{U}\vv v = \lambda^{-1} \boldsymbol{U}\vv v \]
and we can see that for each eigenvector $\vv v$ with eigenvalue of $\lambda$, there must also be an eigenvector $\boldsymbol{U}\vv v$ with eigenvalue of $1/\lambda$.

This result forces the one-cavity system to be reciprocal: swapping the direction of propagation of all the waves interchanges the eigenvalues, with the modulus of the associated propagation constants unmodified.  As expected, this result does not change even if we add free space around the cavity. Considering the ``propagator matrix'' 
\begin{equation*}
    \boldsymbol{P}(L) = \mqty({\E^{\I\vv Kx}} & \\ & {\E^{-\I\vv Kx}}),
\end{equation*} 
we see that it also satisfies $\boldsymbol{P}(L)^{-1} = \boldsymbol{U}\boldsymbol{P}(L)\boldsymbol{U}$, so that the full transfer matrix still satisfies $\boldsymbol{T}^{-1} = \boldsymbol{U}\boldsymbol{T}\boldsymbol{U}$.

\begin{figure}
\begin{center}
	\includegraphics{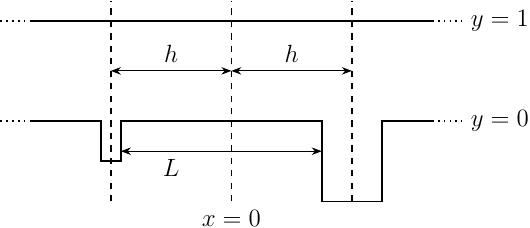}
\end{center}
 \caption{Non-reciprocal geometry.}
\label{fig:appendC}
\end{figure}

As seen above, a system with one cavity will always be reciprocal. To break this constraint we need at least two cavities. This can be illustrated by performing the same steps as in Section~\ref{subsec:b}\ref{subsec:matching}, only considering a waveguide with two cavities as in Fig.~\ref{fig:appendC}. The widths of the cavities will be $a_{1,2}$, the depths $d_{1,2}$. We then see that $h = (L+a_{1}+a_{2})/2$.

Here we outline the calculations. Let us denote the integral in the solution \eqref{eq:phi-vec} by $I$:
\begin{equation}
\boldsymbol{I}(a,x,y) := \lim_{\eta\to0}\int_{-\infty}^{\infty}\frac{{\rm d}k}{2\pi}{\rm e}^{{\rm i}kx}\frac{\sin(ka)}{k}\frac{\cos(\sqrt{(\boldsymbol{K}+{\rm i}\eta\mathbbm{1})^2-k^2\mathbbm{1}}(y-1))}{\sqrt{(\boldsymbol{K}+{\rm i}\eta\mathbbm{1})^2-k^2\mathbbm{1}}\sin(\sqrt{(\boldsymbol{K}+{\rm i}\eta\mathbbm{1})^2-k^2\mathbbm{1}})};
\end{equation}
The result can be written as a superposition of contributions of the both cavities. We obtain an analogue to \eqref{eq:phi-vec}:
\begin{equation}
\phi = \boldsymbol{I}(a_1,x+h,y) \pdv{\boldsymbol{B_1}}{d_1} \boldsymbol{B_1}^{-1} \boldsymbol{\phi}_{c,1} + \boldsymbol{I}(a_2,x-h,y) \pdv{\boldsymbol{B_2}}{d_2} \boldsymbol{B_2}^{-1} \boldsymbol{\phi}_{c,2} + {\rm e}^{{\rm i} \boldsymbol Kx} \boldsymbol{c}^{(+)} + {\rm e}^{-{\rm i} \boldsymbol Kx} \boldsymbol{c}^{(-)}
\end{equation}
where $\boldsymbol B_{1,2}$ are the matrices \eqref{denote-B} written for each cavity separately, and $\boldsymbol{\phi}_{c,1,2}$ are the amplitudes of the three frequencies for both cavities. We can join these two vectors of amplitudes to make one vector $\boldsymbol{\phi}_c$ of 6 entries. Then the previous equation can be written as
\begin{equation}
\phi = \left[\boldsymbol{I}(a_1,x+h,y) \pdv{\boldsymbol{B_1}}{d_1} \boldsymbol{B_1}^{-1} \mqty(\mathbbm{1} & \mathbb{0}) + \boldsymbol{I}(a_2,x-h,y) \pdv{\boldsymbol{B_2}}{d_2} \boldsymbol{B_2}^{-1} \mqty(\mathbb{0} & \mathbbm{1}) \right] \boldsymbol{\phi}_c 
+ {\rm e}^{{\rm i} \boldsymbol Kx} \boldsymbol{c}^{(+)} + {\rm e}^{-{\rm i} \boldsymbol Kx} \boldsymbol{c}^{(-)} \label{appC:phi}
\end{equation}
Now we must match this field to the fields in both cavities. We average over each aperture in turn to obtain:
\begin{align}
    \left[a_1 \boldsymbol{M}(a_1) \pdv{\boldsymbol{B_1}}{d_1} \boldsymbol{B}_1^{-1} \mqty(\mathbbm{1} & \mathbb{0}) + a_2 \boldsymbol{N}_- \pdv{\boldsymbol B_2}{d_2} \boldsymbol{B}_2^{-1} \mqty(\mathbb{0} & \mathbbm{1})\right] \boldsymbol{\phi}_c
    \shoveright{+ \mqty({\rm e}^{-{\rm i} \boldsymbol{K} h} \frac{\sin \boldsymbol{K} a_1}{\boldsymbol{K} a_1} & {\rm e}^{{\rm i} \boldsymbol{K} h} \frac{\sin \boldsymbol{K} a_1}{\boldsymbol{K} a_1}) \boldsymbol{c} = \mqty(\mathbbm{1} & \mathbb{0}) \boldsymbol{\phi}_c,} \\
    \shoveleft{\left[a_1 \boldsymbol{N}_+ \pdv{\boldsymbol{B_1}}{d_1} \boldsymbol{B}_1^{-1} \mqty(\mathbbm{1} & \mathbb{0}) + a_2 \boldsymbol{M}(a_2) \pdv{\boldsymbol B_2}{d_2} \boldsymbol{B}_2^{-1} \mqty(\mathbb{0} & \mathbbm{1})\right] \boldsymbol{\phi}_c}
    \shoveright{+ \mqty({\rm e}^{{\rm i} \boldsymbol{K} h} \frac{\sin \boldsymbol{K} a_2}{\boldsymbol{K} a_2} & {\rm e}^{-{\rm i} \boldsymbol{K} h} \frac{\sin \boldsymbol{K} a_2}{\boldsymbol{K} a_2}) \boldsymbol{c} = \mqty(\mathbb{0} & \mathbbm{1}) \boldsymbol{\phi}_c,} \\
    \label{appC:averages}
\end{align}
where $\boldsymbol{M}(a)$ is the overlap integral \eqref{eq:M-vector} of each cavity with itself, and $\boldsymbol{N}_\pm$ are overlap integrals of the cavities with each other, defined as
\begin{equation}
    \boldsymbol{N}_\pm := \frac{1}{\pi} \lim_{\eta\to0^+} \int_{-\infty}^\infty \frac{\sin ka_1}{ka_1} \frac{\sin ka_2}{ka_2} \E^{\pm2\I kh} \frac{\cot \sqrt{(\boldsymbol K+\I\eta\mathbbm{1})^2 - k^2\mathbbm{1}}}{\sqrt{(\boldsymbol K+\I\eta\mathbbm{1}})^2 - k^2\mathbbm{1}}\,\D k,
\end{equation}

Now the two equations \eqref{appC:averages} can be merged into one $6\times6$ matrix equation which can be used to express $\boldsymbol{\phi}_c$ in terms of $\boldsymbol{c}$ as
\[ \boldsymbol{\phi}_c = \boldsymbol{R} \boldsymbol{c},\]
where
\[ \boldsymbol{R} := - \left[\mqty(a_1 \boldsymbol{M}(a_1) \pdv{\boldsymbol{B}_1}{d_1} \boldsymbol{B}_1^{-1} & a_2 \boldsymbol{N}_- \pdv{\boldsymbol B_2}{d_2} \boldsymbol{B}_2^{-1} \\ a_1 \boldsymbol{N}_+ \pdv{\boldsymbol{B_1}}{d_1} \boldsymbol{B}_1^{-1} & a_2 \boldsymbol{M}(a_2) \pdv{\boldsymbol B_2}{d_2} \boldsymbol{B}_2^{-1}) - \mathbbm{1} \right]^{-1} \mqty(\E^{-\I\vv Kh} \frac{\sin \vv Ka_1}{\vv Ka_1} & \E^{\I\vv Kh} \frac{\sin \vv Ka_1}{\vv Ka_1} \\ \E^{\I\vv Kh} \frac{\sin \vv Ka_2}{\vv Ka_2} & \E^{-\I\vv Kh} \frac{\sin \vv Ka_2}{\vv Ka_2}), \]
which, in turn, can be plugged into \eqref{appC:phi} to obtain
\begin{equation}
\boldsymbol\phi = \left\{ \left[\boldsymbol{I}(a_1,x+h,y) \pdv{\boldsymbol{B_1}}{d_1} \boldsymbol{B_1}^{-1} \mqty(\mathbbm{1} & \mathbb{0}) + \boldsymbol{I}(a_2,x-h,y) \pdv{\boldsymbol{B_2}}{d_2} \boldsymbol{B_2}^{-1} \mqty(\mathbb{0} & \mathbbm{1}) \right] \boldsymbol{R} 
+ \mqty({\rm e}^{{\rm i} \boldsymbol Kx} & {\rm e}^{-{\rm i} \boldsymbol Kx}) \right\}\boldsymbol{c}. \label{appC:solution}
\end{equation}
This can be used for obtaining the transfer matrix of a pair of cavities. If we once more introduce the far-field approximation $\boldsymbol{I} \approx \frac{\sin \boldsymbol{K} a}{\boldsymbol{K}} \frac{\E^{\I\boldsymbol{K}|x|}}{\I\boldsymbol{K}}$ and relating the fields on the far left of the pair of cavities with the fields on the far right, we can obtain the transfer matrix in the following form:
\begin{equation}
    \vv T =  \left[\mathbbm{1}_{6\times6} - \mqty(\boldsymbol Q_+ \\ \mathbb{0}_{3\times6})\right] \left[ \mathbbm{1}_{6\times6} - \mqty(\mathbb{0}_{3\times6} \\ \boldsymbol Q_-) \right]^{-1}, \label{appC:T}
\end{equation} 
where $\mathbbm{1}_{m\times n}$ and $\mathbb{0}_{m\times n}$ denote identity and zero matrices of given size, and 
\begin{align}
\boldsymbol Q_\pm := \I \bigg[ a_1 \frac{\sin\vv Ka_1}{\vv Ka_1} \frac{\E^{\pm\I\vv K h}}{\vv K} \pdv{\vv B_1}{d_1} \vv B_1^{-1} \mqty(\mathbb 1 & \mathbb 0) + a_2 \frac{\sin\vv Ka_2}{\vv Ka_2} \frac{\E^{\mp\I\vv K h}}{\vv K} \pdv{\vv B_2}{d} \vv B_2^{-1} \mqty(\mathbb 0 & \mathbb 1) \bigg] \boldsymbol R.
\end{align}

The transfer matrix in \eqref{appC:T} is manifestly non-reciprocal, which means that a generic two-cavity setup will exhibit non-reciprocity. The advantage of \eqref{appC:T} over a simple product of three transfer matrices of the form \eqref{eq:T-cav-final} lies in the fact that the calculation just shown accurately accounts for the interaction of the two cavities (while each transfer matrix \eqref{eq:T-cav-final} by itself introduces a far-field approximation, so their product only accounts for the interaction of the fundamental modes of both cavities).


%

\end{document}